# Low Frequency Oscillations of Plasma Injected in Open Magnetic Trap from Independent Ultra-high Frequency Source

S. Nanobashvili[1], Z. Beria[1], G. Gelashvili[1], G. Gogiashvili[1],
I. Nanobashvili[1], G. Tavkhelidze[1], G. Van Oost[2]

[1]*Andronikashvili Institute of Physics of the Iv. Javakhishvili Tbilisi State University, Georgia*
[2]*Department of Applied Physics, Ghent University, Belgium*

Methods of plasma heating by electromagnetic waves, interaction of electro-magnetic waves with magnetoactive plasma, plasma turbulence and transport processes are investigated in the trap of the linear plasma device – OMT-2. An ultrahigh frequency (UHF) contactless method is used to fill the open magnetic trap by plasma. We propose a new method of open magnetic trap filling with plasma whereby plasma is injected in the trap along the magnetic field from an independent stationary UHF source. The source is located outside the trap and plasma formation in it takes place in the strongly nonuniform magnetic field, in the electron cyclotron resonance (ECR) regime by means of the UHF power. In this paper we study the efficiency of open magnetic trap filling by plasma and plasma low frequency (LF) oscillations in the trap.

## 1. Introduction

Different methods of filling open magnetic trap with plasma are used in various experiments. Contactless methods are used most frequently and lately UHF methods of plasma accumulation in a trap were most widely used. As a rule, plasma formation takes place in the trap itself in ECR regime (see e.g. [1-3]). However, this method has various disadvantages. In particular, the range of magnetic field variation in the trap is strictly limited by the existence of UHF discharge in the magnetic field. Together with change of magnetic field discharge regime and plasma parameters change. The most important disadvantage is that the "hot" region of UHF wave absorption in the plasma is in the trap itself, which is often undesirable. Therefore, the application of an independent plasma source with controllable parameters located far from the trap and from which the "target" plasma is injected in the trap is of great interest. The present work deals with this new application. The independent stationary UHF plasma source is described together with its characteristics. Possibility of filling an open magnetic trap with uniform field by plasma injected from the source as well as the properties of plasma and its LF oscillation characteristics in the trap are investigated.

## 2. Experimental set-up

Experiments were carried out on a stationary installation OMT-2 (Fig.1,2).

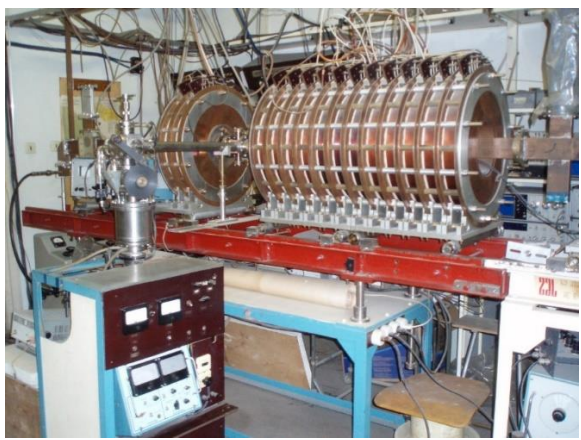 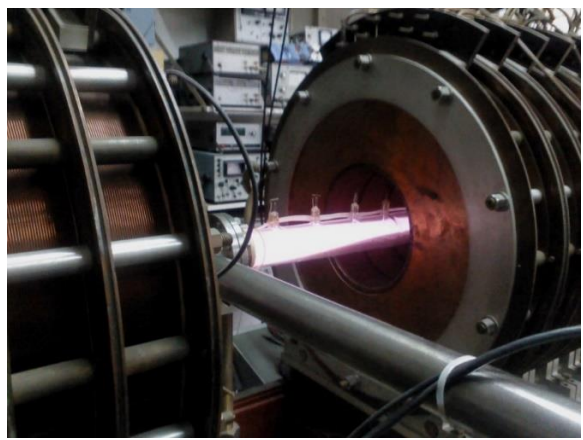

**Fig.1** Installation OMT-2       **Fig.2** Plasma in OMT-2

The diagram of OMT-2 is presented in Fig.3. It consists of two main parts: an independent UHF plasma source and open magnetic trap, in which plasma is injected.

OMT-2 plasma device is open magnetic trap with length 90 cm and inner diameter 19.5 cm. Stationary magnetic field in the trap is produced by solenoid composed of 12 identical coils. Each coil consists of 4 windings. Depending on the way of winding connection inside coils and coil connection with each other it is possible to obtain by means of solenoid – uniform magnetic field with length 45 cm and uniformity on the axis of solenoid of the order 0.1 %, magnetic field of mirror and multimirror configuration with controllable mirror ratio and field with opposite magnetic fields (configuration picket fence). Feeding of solenoid is done

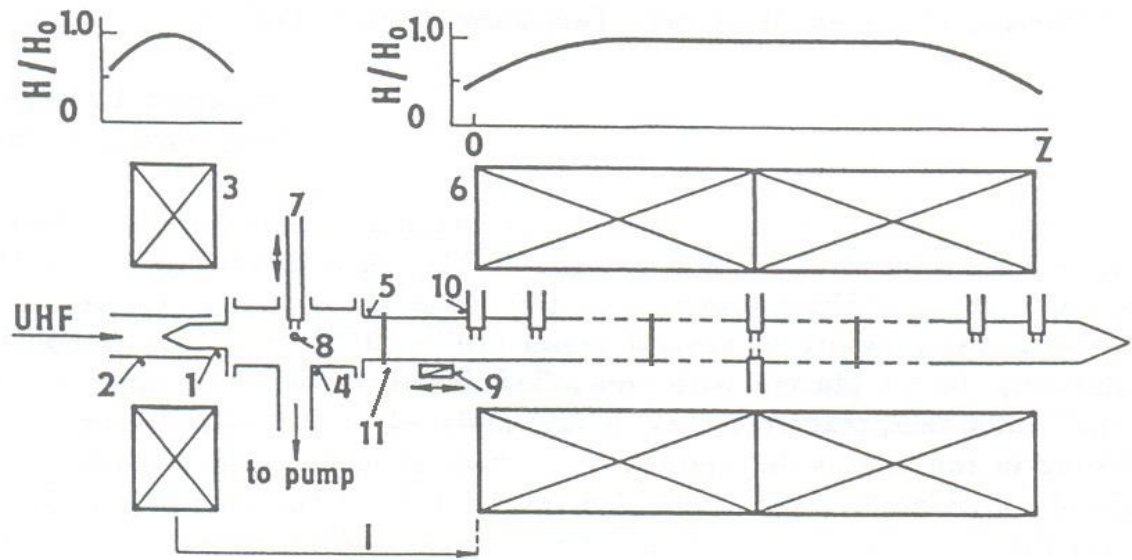

**Fig.3** The scheme of experimental set-up

*1* – Discharge chamber, *2* – rectangular waveguide, *3* – coil forming the magnetic field in source, *4* – diagnostic section, *5* – volume under investigation, *6* – solenoid, *7,10* – double electric probes, *8,9* – semiconducting light sensors, *11* – capacitance probes by means of amplidynes. Magnetic field strength in the solenoid can be varied smoothly from 0 to its maximum value by variation of current from amplidynes.

Maximum uniform magnetic field on the axis of solenoid reaches 5 kOe. At maximum value of magnetic field in the solenoid electric power consumed by amplidynes equals to 30 kW.

Stationary magnetic field in independent UHF source of plasma is created by means of 2 coils (complete length 15 cm). These coils are identical to the coils of magnetic trap solenoid. Stationary magnetic field with maximum at the centers of coils is produced by means of amplidyne. Magnetic field can be varied smoothly from 0 to the maximum value 2 kOe.

In Fig. 1 above solenoid and coils of independent UHF plasma source qualitative picture of magnetic field distribution is shown in solenoid (in case of uniform field) and also in plasma source. For each case $H_0$ is corresponding maximum value of magnetic field in the center and H – magnetic field along *z* axis.

In the UHF source plasma is formed in a quartz tube *(1)* with inner diameter 2.6 cm and length 10 cm. Stationary UHF power is supplied to the discharge chamber by a standard rectangular waveguide *(2)* in which $TE_{01}$ is excited. As a stationary UHF generator we use stationary magnetron at the frequency 2400 MHz. UHF power can be varied smoothly in the range 0÷150 W by stabilized current variation in the magnetron and also by attenuator in the feeding waveguide tract of independent UHF plasma source.

In the independent UHF plasma source the discharge chamber *(1)* and the waveguide *(2)* is in the stationary magnetic field, created by a short coil *(3)*.

The discharge chamber of the UHF plasma source is connected with the cylindrical section *(4)*, made of stainless steel serving as a diagnostic section [4].

The investigation volume *(5)* with plasma is placed in the stationary magnetic field, created by a solenoid *(6)*.

The results presented in the this paper deal with the plasma injection into the trap with uniform magnetic field.

The distance between UHF plasma source and main investigation volume (the trap) can be varied in the range $\ell = 30 \div 90$ cm. In this experiments $\ell = 45$ cm. In the described experiments a glass tube with inner diameter of 6 cm and length 90 cm has been used as investigation volume *(5)*. The conditions of discharge existence in the UHF plasma source where determined using semiconducting light sensor *(8)* which records the plasma integral light emission. The injected plasma parameters – density of charged particles, temperature of electrons and their distribution over the radius - were measured by a movable double electric probe *(7)*.

In order to determine the efficiency of the magnetic trap filling with plasma and study its characteristics in the trap (density and temperature of electrons) semiconducting light sensor *(9)* moving along the chamber *(5)* and double electric probes *(10)* inserted along the chamber axis in 8 locations with 7.5 cm step were used. In our case the plasma is weakly ionized. As it is well known [5] integral emission of plasma depends on its particle density. We have verified this experimentally. At the same time, electric probes can also be used for local determination of the spectrum of plasma oscillations. Plasma oscillations in the trap are also detected by capacitance probe *(11)*. The capacitance probe is a ring of copper foil with width 5 mm and non-connected ends. It registers variation of potential on the capacitor plasma-ring when oscillations of plasma parameters exist in this region. Oscillation spectrum is registered by means of low-frequency spectrum analyzer.

During the experiments described above Argon and Helium were used as working gas. Plasma in the UHF source was created at working gas pressure $10^{-5} \div 10^{-2}$ Torr.

## 3. Results and discussion

**1) UHF plasma source.** Discharge in the source can be obtained in the investigated range of working gas pressure by means of stationary UHF power only if the condition of electron cyclotron resonance is fulfilled in the region of UHF field interaction with plasma, $\omega_0 = \omega_{He} = eH/mc$, i.e. when the magnetic field equals to that of the cyclotron one ($H_c = 850$ Oe).

After the appearance of a discharge plasma exists even under the change of magnetic field in certain limits depending on the pressure of neutral gas and UHF power supplied to plasma. As experiments show, at low gas pressure plasma exists at magnetic fields not significantly different from cyclotron field. With increase of pressure the region of discharge existence broadens significantly towards the magnetic fields lower than cyclotron. This is in good agreement with the results of investigations of UHF discharge in the magnetic field (see e.g. [1]). In this region UHF power absorption is determined by linear transformation of the wave in the upper hybrid resonance. The measurements by double electric probe *(7)* have shown that by changing supplied UHF power, neutral gas pressure and magnetic field of coil *(3),* one can change in wide range the density of plasma injected from the source into the trap.

Under our experimental conditions plasma density can be changed from $10^8$ to $10^{12}$ cm$^{-3}$ with plasma electron temperature $T_e = 2 - 3$ eV. These changes are quite controllable and well reproducible.

**2) Plasma injection in magnetic trap.** Plasma with above given parameters has been injected from UHF source into uniform magnetic field and magnetic trap of mirror configuration formed by a solenoid *(6)* (Fig.3). Filling of open magnetic trap with plasma has been studied experimentally in a wide range of parameters of the injected plasma, neutral gas pressure, magnetic field strength in the trap and also the distance between plasma source and solenoid.

Typical results of experimental determination of the efficiency of open magnetic trap filling by plasma – dependence of the parameter $I/I_0$ on *z*, under different pressure, measured by movable light sensor *(9)*, are presented in the Fig.4, where $I_0$ is the maximum value of light emission intensity of plasma at the entrance of the magnetic trap (solenoid), $I$ - light emission intensity of plasma in magnetic trap and $z$ – coordinate along the axis of the solenoid. In Fig.5 the dependence of the parameter $I/I_0$ on $z$ are presented for different values of magnetic field in the center of the trap.

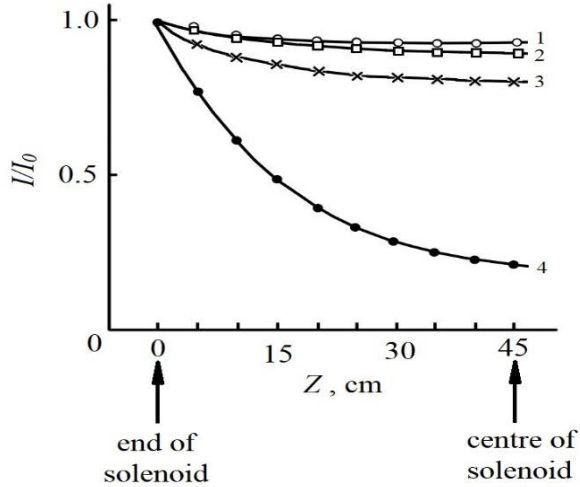

**Fig.4**
*1 - $p=2.3 \cdot 10^{-4}$ Torr; 2 - $p=4.6 \cdot 10^{-4}$ Torr; 3 - $p=7.6 \cdot 10^{-4}$ Torr; 4 - $p=2.3 \cdot 10^{-3}$ Torr; for **$H_t=400$ Oe***

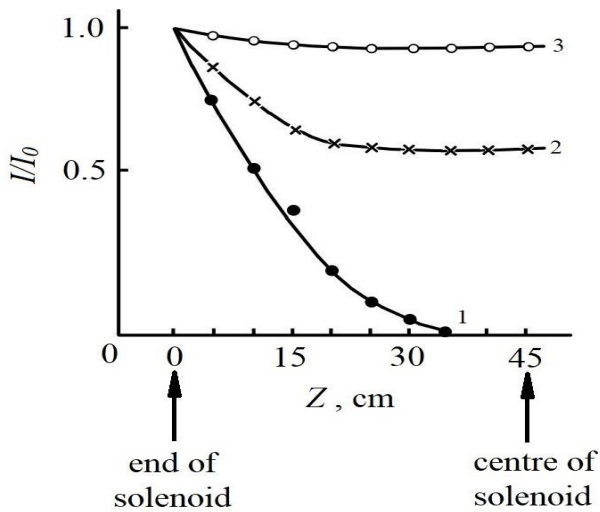

**Fig.5**
*1 - $H_t=0$; 2 - $H_t=100$ Oe; 3 - $H_t=400$ Oe; for **$p=2.3 \cdot 10^{-4}$ Torr***

Experimental data analysis allows to conclude – for pressure $p < 1 \cdot 10^{-3}$ Torr, magnetic field in the trap $H_t > 400$ Oe and distance between the plasma source and solenoid $\ell < 80$ cm, the filling of the trap is very efficient and plasma with controllable density within the range $10^8 \div 10^{12}$ cm$^{-3}$ and temperature $2 - 3$ eV is accumulated in the trap.

**3) Low frequency oscillations of plasma in magnetic trap.** Experiments showed that in the whole investigated range of conditions, oscillations in the low frequency (LF) range of spectrum (measured by capacitance probe) are excited in the plasma located in the magnetic trap. These oscillations are excited more efficiently at certain "critical" values of magnetic field in the region of UHF power absorption in the plasma source, in particular, at electron cyclotron resonance and its harmonics, i.e. when $\omega_0 = N\omega_{He}$.

The picture in Fig.6 shows in arbitrary units a typical plasma oscillation spectrum in the low frequency range.

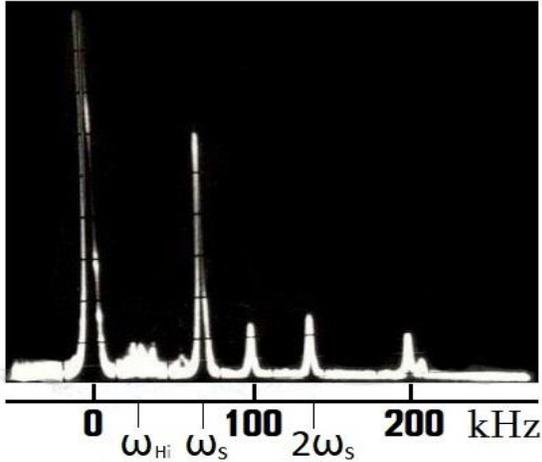

**Fig.6** Plasma oscillation

In Argon plasma we detect oscillations with frequency ~70 kHz, which practically does not change with modification of the magnetic field neither in the plasma source, nor in the magnetic trap. Besides, oscillations with frequency ~30 kHz are observed in the plasma and this frequency changes proportionally to the magnetic field.

The amplitude of the LF oscillations increases with UHF power injected in the plasma and when $P_{UHF} > 50W$ second and even third harmonic of LF oscillations with frequency ~140 kHz and ~210 kHz appear (Fig. 6). As for the oscillations with frequency ~30 kHz, excitation of high harmonics is not observed even with maximum of UHF power injected in plasma $P_{UHF} = 150W$.

It must be mentioned that LF oscillations of plasma are observed down to minimum UHF power $P_{UHF} = 5W$ injected into plasma for which it is still possible to maintain a discharge in plasma source.

It is known [6-9] that in non-equilibrium magnetoactive plasmas where the condition $T_e \gg T_i$ is fulfilled ion cyclotron oscillations with frequency

$$\omega_{Hi} = \frac{ZeH}{M_i c} \quad (1)$$

($H$ is magnetic field strength, $Z$ - number of ion charge and $M_i$ - ion mass) and ion acoustic oscillations with frequency

$$\omega_s = \frac{\pi\, l}{L} \cdot \left(\frac{3kT_e}{M_i}\right)^{1/2} \qquad (2)$$

($l$ is oscillation mode number, $L$ - characteristic size of plasma and $T_e$ - electron temperature) may appear.

Estimations show that oscillations observed at a frequency of the order of 30 kHz under our experimental conditions correspond to ion cyclotron oscillations (1) in terms of both – frequency and character of dependence on magnetic field.

Concerning oscillations at a frequency of the order of 70 kHz, calculations according to the (2) for our experimental conditions ($T_e$ = 2 – 3 eV, $L$ = diameter of discharge chamber and $l = 1$) correspond well to the value of experimentally measured ion sound waves with wavelength equal to the inner diameter of the discharge chamber.

Since both ion cyclotron (1) and ion sound (2) frequency depend on ion mass, changing the working gas should cause a shift of the frequency of the LF oscillations. Indeed, dedicated experiments in Helium showed that the frequency of the LF oscillations shifts according to the modification of mass. This is an additional argument in support of the statement that under our experimental conditions excitation of ion sound and ion cyclotron oscillations takes place in plasma.

## 4. Conclusion

The presented experimental results allow to conclude that:

a) the proposed method of magnetic trap filling with plasma allows to accumulate plasma with controllable parameters in an open magnetic trap

b) under the above mentioned expermental conditions near the UHF plasma source and also in the trap first and second harmonics of ion sound wave together with ion cyclotron oscillations are reliably detected.

**References**

[1] Anisimov A.I., Budnikov V.N., Vinogradov N.I., Golant V.E., Nanobashvili S.I., Obukhov A.A., Pakhomov L.P., Pilya A.D., Fedorov V.I. Study of UHF Plasma Heating in Magnetic Field // Plasma Phys. And Contr. Nucl. Fus. Res., IAEA, Vienna, Part II, p.399, 1969.

[2] Zalesskii I.G., Komarov A.D., Lavrentyev O.A., Naboka V.A., Nazarov N.I., Potarenko V.A., Stepanenko I.A. Plasma Production in an Electrostatically Plugged Mirror System by Microwave Method // Sov. J. of Plasma Phys., [Fizika Plazmy], **5,** 532, 1979.

[3] Nanobashvili S.I., Datlov J., Stockel J., Zacek F. Plasma Formation and sustainement by a Multijunction Grill on the CASTOR Tokamak // Czech. J. Phys., **B37**, 194, 1987.

[4] Gogiashvili G.E., Nanobashvili S.I., Rostomashvili G.I. Stationary UHF Plasma Source // Sov. J. Tech. Phys., [JTF], **57**, 1746, 1987.

[5] Golant V.E. Diffusion of Charged Particles in a Plasma in a Magnetic Field // Sov. Physics Uspekhi. **6**, 161, 1963.

[6] Silin V.P., Rukhadze A.A. Electromagnetic Properties of Plasma and Plasmalike Media // 3rd Edition, Publishing House Librokom, Moscow, 2013.

[7] Ginzburg V.L., Rukhadze A.A. Waves in Magnetoactive Plasma, // 2nd Revised Edition, Izdatel'stvo: Nauka, Moskva, 1975. (In Russian).

[8] Akhiezer A.I., Akhiezer I.A., Polovin P.B., Sitenko A.G., Stepanov K.N. Collective Oscillation in Plasma // Izdatel'stvo: Atomizdat, Moskva, 1964. (In Russian).

[9] Alexandrov A.F, Bogdankevich L.S., Rukhadze A.A. Principles of Plasma Electrodynamics // Springer Series-Plasma Physics, 1984.